Perspective

# Probing laser-driven structure formation at extreme scales in space and time

Jörn Bonse[1] and Klaus Sokolowski-Tinten[2]

[1] Bundesanstalt für Materialforschung und -prüfung (BAM), Berlin, Germany
[2] Faculty of Physics and Centre for Nanointegration Duisburg-Essen, University of Duisburg-Essen, Duisburg, Germany

E-mail: joern.bonse@bam.de , klaus.sokolowski-tinten@uni-due.de

**Abstract**

Irradiation of solid surfaces with high intensity, ultrashort laser pulses triggers a variety of secondary processes that can lead to the formation of transient and permanent structures over large range of length scales from mm down to the nano-range. One of the most prominent examples are LIPSS – Laser Induced Periodic Surface Structures. While LIPSS have been a scientific evergreen for of almost 60 years, experimental methods that combine ultrafast temporal with the required nm spatial resolution have become available only recently with the advent of short pulse, short wavelength free electron lasers. Here we discuss the current status and future perspectives in this field by exploiting the unique possibilities of these 4$^{th}$-generation light sources to address by time-domain experimental techniques the fundamental LIPSS-question, namely why and how laser-irradiation can initiate the transition of a "chaotic" (rough) surface from an aperiodic into a periodic structure.

Keywords: Free electron laser, Time-resolved scattering,·Pump-probe experiments,·Laser-induced periodic surface structures (LIPSS), Capillary waves, Theoretical modelling·

---

## 1. Introduction

Irradition of solid surfaces with intense ultrashort laser pulses allows to create high energy density, non-equlibrium states of condensed matter. Subsequent to the initial deposition of energy into the electronic system a complex chain of secondary relaxation processes can lead to rapid structural changes, often along unusual, non-equilibrium pathways. Depending on the specific material properties and the excitation strength this may include sub-ps electron thermalization, phonon non-equilibrium and ultrafast lattice heating, excitation of coherent phonons, thermal and non-thermal melting, hydrodynamic melt flows, ablation by spallation, phase explosion and overcritical expansion, rapid resolidification by crystallization or amorphization, liquid-liquid phase transitions, as well as chemical reactions with the ambient environment (e.g. [1 – 21] – a by far incomplete and most likely "biased" selection).

*1.1 Multi-scale nature of laser ablation: atomic-, meso-, and micro-scale*

The complete chain of all these processes is intrinsically of multiscale nature, both in time and space [22 - 24]. It starts with processes acting almost instantaneously upon laser irradiation, i.e. within just a few femtoseconds, while reaching the final state can take up to the millisecond range, thus covering more than 10 orders of magnitude in time. The corresponding spatial dimensions are ranging from interatomic distances (~0.1 nm), over meso- and micro-scale patterns towards the millimeter regime, covering again many orders of magnitude. As an outcome of these complex spatio-temporal multi-scale processes, the irradiation of solids with intense laser pulses results in a large variety of structures on the irradiated surfaces down to the nano-scale.

*1.2 Laser-induced periodic nanostructures (LIPSS): a universal phenomenon*

Among them the so-called *laser-induced periodic surface structures* (LIPSS) are one of the most prominent examples [25-30]. These structures represent a universal phenomenon [25] and can already be considered as a scientific evergreen [31] being studied for more than five decades after their first experimental observation in 1965 [32].

LIPSS manifest as 1D or 2D quasi-periodic gratings imprinted either in the surface topography or in other material surface properties. They feature either spatial periods of the order of the laser irradiation wavelength λ (so-called *low spatial frequency LIPSS*: LSFL) or even significantly smaller (as small as a few tens of nanometers; *high spatial frequency LIPSS*: HSFL).

It is now generally accepted that LSFL result from the interference of the incident laser wave with waves scattered/excited at the surface leading to a (periodically) modulated energy deposition [33]. Already in 1983 Sipe et al. [34] developed this qualitative picture into a rigorous electromagnetic model which is frequently used and has been further developed since then [35 - 37]. The analytical Sipe model represents a static approach that considers the absorption of electromagnetic radiation in a microscopically rough near-surface region with fixed optical properties. As such, it does not account for material-specific transient responses

(e.g. changes of the optical properties, thermophysical effects, etc.) that can lead to intra- or multi-pulse feedback effects and may significantly alter the LIPSS-formation process. Moreover, as result of some inherent mathematical approximations, the applicability of Sipe's theory may be limited for very small structures such as HSFL.

1.3 Falling below the diffraction limit

HSFL with a periodicity much smaller than the laser wavelength (see Fig. 1) have been observed only after irradiation with fs- and ps-laser pulses. Two different types of HSFL are empirically distinguished: "Deep" HSFL-I with spatial periods of a few hundreds of nanometers and a depth-to-period aspect ratio $A > 1$ are formed on dielectrics for sub-bandgap excitation [Fig. 1(a), [38]], while very "shallow" ($A << 1$) HSFL-II with spatial periods less than 100 nm are observed on some metals [Fig. 1(b), [39]].

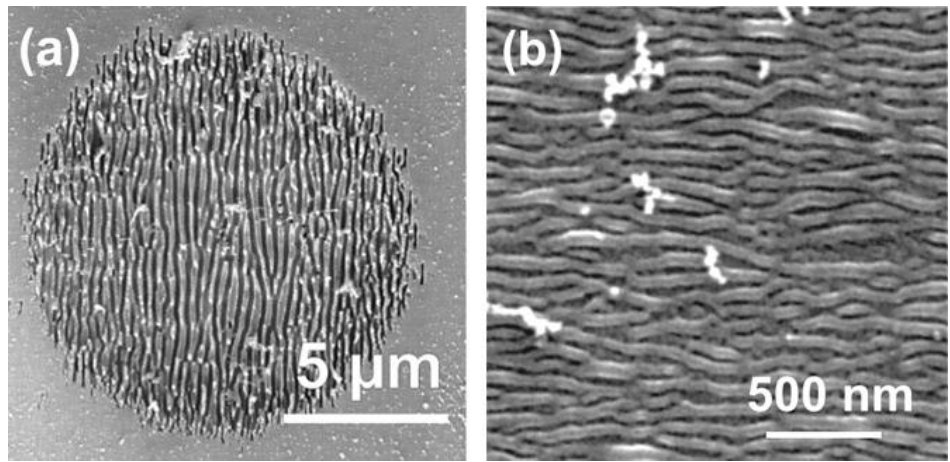

**Fig. 1:** SEM micrographs of HSFL-I on $SiO_2$ (a) and HSFL-II (b) on Ti surfaces after irradiation with multiple fs-laser pulses in air (pulse duration 30 fs / 120 fs, wavelength ~800 nm) [38, 39]. Note the different magnifications. (a) Reproduced from [38] under the terms of a Creative Commons BY 4.0 license. Copyright 2017, The Authors, published by Springer Nature. (b) Reproduced from [39], Bonse J, et al., Appl. Phys. A 110 (2013) 547, Springer-Verlag, Copyright 2012, reproduced with permission from Springer Nature.

The mechanisms leading to these very different forms of HSFL are not fully clear yet [40]. The lower limit of HSFL periods mentioned above may be determined by fundamental energy-relaxation processes following the initial non-equilibrium excitation and the typical spatial decay lengths of optical near-fields in the vicinity of locally excited scattering centers [41, 42].

*1.4 Surface functionalization through LIPSS*

LIPSS currently gain a lot of attention since they can be generated on almost any material and allow to functionalize surfaces with a great potential for technical applications. They can be processed on large surface areas in a straightforward and reliable manner, being compatible with industrial demands and technologies [30]. Given their (deep) sub-micrometric size, such surface nanostructures can act as optical diffraction gratings featuring structural colors [43] or can impose sub-wavelength scattering effects [44, 45] allowing to manage and control optical surface properties. Moreover, these surface nanostructures can affect other mechanical, structural, or electronic properties, such as the coefficient of friction, the surface wettability, enhancing locally electromagnetic fields, they are stimulating or suppressing cell growth, or can reduce biofilm formation, thus, enabling applications in the fields of tribology, mechanical engineering, biology, plasmonics, photovoltaics, and medicine [30, 46 - 49].

*1.5 Time-resolved studies of LIPSS: current limits*

In the vast majority of published work LIPSS-formation is addressed rather indirectly through the *post mortem* analysis of the permanent modifications at the irradiated surface. In contrast, only a comparatively small number of time-resolved experiments have been performed. Due to the inherent multiscale nature of these irreversible processes – both temporally and spatially – the direct observation of LIPSS formation has remained challenging, particularly for sub-wavelength structures.

The potential of using light that is being diffracted or scattered at the developing periodic surface topography of LIPSS as a probe for the formation dynamics was already recognized during the early eighties of the past century [50 - 52]. At that time, ns-pulsed lasers were used to generate LIPSS, while simultaneously probing the pattern of scattered or diffracted light with an additional continuous wave (cw) laser. In these experiments, the available pulse duration in the ns-range and intrinsic photodetector response times limited the temporal resolution. Moreover, with such long pulses, the coupling of optical energy into the material occurs quasi simultaneously with its structural response, i.e. heating, phase transitions and ablation already take place during irradiation and are temporally interleaved.

This changed with the availability of ultrashort pulsed lasers with pulse duration in the few ps to fs range. They enable excitation faster than the typical electron-phonon relaxation times, thus, seperating energy deposition and the subsequent material response in the time domain. Furthermore, short pulses allowed to explore the dynamics of LIPSS-formation with high temporal resolution even on ultrashort timescales via so-called *pump-probe* experiments. In such experiments, an ultrashort optical laser pump-pulse excites the surface, while a second temporally delayed optical probe-pulse interrogates the laser-excited state of matter at a chosen delay time, acquiring its signature with a *slow* detector (e.g. photodiode, CCD). Recording such snapshots as a sequence of different delay times allows to track the temporal evolution of the surface structures in a stroboscopic fashion with a time-resolution given by the duration of the probe pulse. Such pump-probe experiments were performed on LIPSS either in optical diffraction geometry [38, 53], in transient optical scattering experiments [54, 55], or even in time-resolved optical bright-field microscopy schemes [56, 57].

As an example we show here results from [53] and [56] that are demonstrating the capabilities and limitations of all-optical time-resolved experiments. In 2013, Höhm et al. presented ultrafast pump-probe experiments using LSFL on fused silica as (transient/permanent) diffraction gratings. The first-order diffraction signal was recorded in transmission geometry at 400 nm probe wavelength, after $N$ pump pulses (50 fs duration, 800 nm wavelength) previously had "dressed" the surface at suitable laser fluences [53]. At a low number of surface dressing pump pulses ($N = 1 - 3$) and even before a permanent surface relief of LSFL could be observed, an ultrafast transient diffraction at the LSFL spatial frequencies was evidenced in the transparency regime of the sample (Figure 2, top part, top row). After the 4-th pump pulse, a permanent LSFL surface relief was formed in the probed spot, accompanied with the excitation of an optically thick laser-induced free electron-plasma at the surface that was dynamically shielding the probe beam (Figure 2, top part, bottom row) for $N > 5$. These effects were attributed to an interplay of a transient refractive index grating formed within < 300 fs by self-trapped excitons (STE's), local heating, plasma relaxation, and ablation.

Three years later, Garcia-Lechuga et al. demonstrated direct in-situ imaging of the dynamics the formation of non-ablative supra-wavelength LSFL on silicon by using "moving-spot" fs-time resolved bright-field microscopy [56], see Fig. 2(a). They kept the peak fluence of a fs-pump beam (120 fs duration, 800 nm wavelength) closely below the ablation threshold fluence, resulting in LSFL quenched from a transient melt pattern into periodic stripes of amorphous material (a-Si) alternating at the crystalline wafer surface. By selecting the suitable sample scanning direction, i.e. when the in-plane wave vector component of the laser radiation is anti-collinear to the sample scanning direction), supra-wavelength-sized amorphous LSFL patterns with periods of ~3.5 µm could be "written" in a continuous line at the laser-excited surface. The synchronized probe beam (400 nm wavelength) then allowed the recording of snapshots of the surface reflectivity at different delay times ranging between a few hundred fs up to several ns (as individually indicated in the micrographs (b) – (e) displayed in the bottom part of Fig. 2).

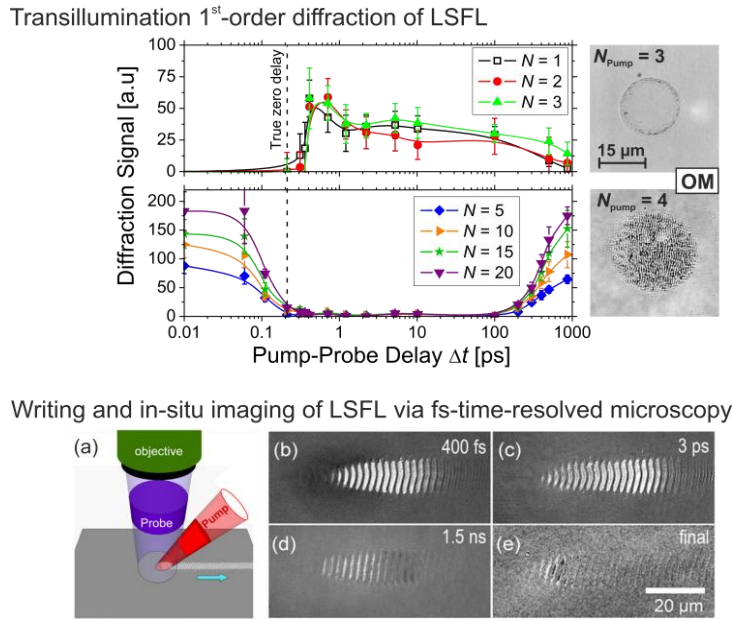

**Fig. 2:** Time-resolved optical pump-probe experiments performed on LIPSS. Top: 1st-order transillumination signal of the probe beam diffracted at the LSFL generated on fused silica by $N$ previous pump pulses (left). Optical micrographs taken after the $N_{pump}$-th pump-pulse are complemented at the right. Reproduced from [53], Höhm S, et al., Appl. Phys. Lett. 102 (2013), 054102, with permission of AIP Publishing. Bottom: Scheme of the continuous (line) writing of supra-wavelength non-ablative LSFL on silicon via laser-induced melting and subsequent amorphization (a), while performing fs-time-resolved bright-field microscopy at different delay times $\Delta t$ = 400 fs (b), 3 ps (c), 1.5 ns (d), and for the final surface state (e). Reprinted (adapted) with permission from [56], Garcia-Lechuga M, et al., ACS Photonics 3 (2016), 1961, Copyright 2016, American Chemical Society.

### 1.6 The need for intense, ultrashort, and short wavelength pulses

All-optical pump-probe techniques are mature and have reached a very sophisticated level. Their spatial resolution, determined by the optical probe wavelength, represents a fundamental limit with respect to the investigation of sub-µm structural changes, such as LIPSS. While near-wavelength sized LSFL are accessible to a certain extent (as discussed above), the formation of all smaller structues, i.e. HSFL, are out of reach. Therefore, to temporally **and** spatially resolve the processes of structure formation on the relevant fs/ps-time- and sub-µm/nm spatial scale the probe pulses must not only be of ultrashort duration, but also exhibit a short wavelength in the XUV or X-ray spectral range – comparable or smaller than the feature sizes that are investigated. Finally, the formation of LIPSS usually occurs in an excitation regime where laser irradiation leads to permanent (irreversible) surface modifications. Hence, the used probe pulses must also be sufficiently intense to enable single-pulse probing of the induced irreversible dynamics.

## 2. Short wavelength FEL sources for probing transient structure formation

While ultrafast, short wavelength pulses, for example through high-harmonic generation in gases or from laser-produced plasmas, had been available for quite some time, these laboratory-scale source usually lack the required photon flux for single-pulse probing of irreversible events like LIPSS-formation with the required signal-to-noise ratio. With the advent of short pulse XUV and X-ray free electron lasers this situation has changed dramatically. These 4th-generation light sources exhibit very high photon flux (up to mJ pulse energies), ultrashort pulse durations of a few tens of fs (or even down to the few-fs and as range) and full spatial coherence. This unique combination properties has opened up completely possibilities, in particular to investigate the dynamic structural response of laser-irradiated matter (e.g. [58, 59] and references therein).

Surprisingly, these possibilities have been so far - with the exception of our own work [60] - not applied to the otherwise intensively studied problem of LIPSS-formation. Therefore, to motivate the future perspectives in this field, as discussed in section 3, we will present in the following a few key results of our proof-of-principle experimens carried out at the FLASH-facility at DESY (Hamburg, Germany), the first FEL operating at wavelengths below 50 nm [61].

*2.1 Probing laser-driven structure formation at extreme spatio-temporal scales in reciprocal space*

The experiments discussed in this section have been enabled through a close collaboration with H. N. Chapman, A. Barty and co-workers using their pioneering experimental platform developed for coherent diffractive imaging experiments [62, 63]. Figure 3 shows a schematic of the optical pump / XUV probe scattering experiment.

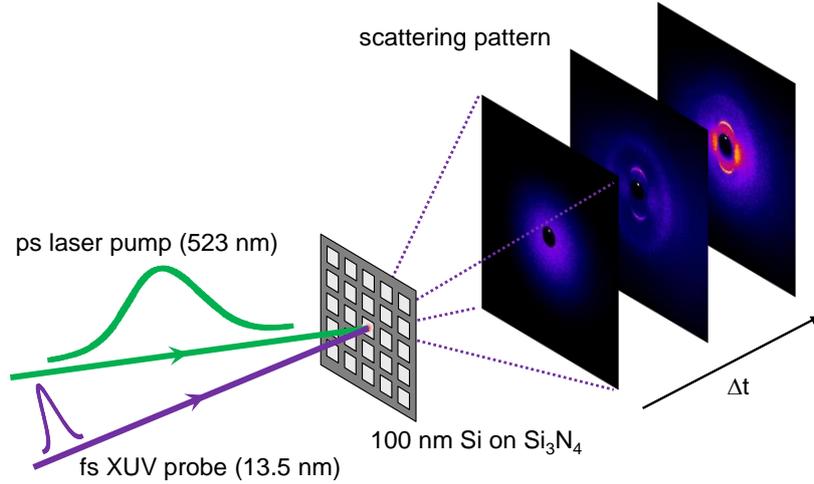

**Fig. 3:** Schematic of the time-resolved single-pulse scattering experiments performed at the XUV-FEL (DESY, Hamburg, Germany).

FLASH was operated in single bunch, ultrashort-pulse mode, delivering pulses with 13.5 nm wavelength, 10 – 20 fs pulse duration, and about 10 – 20 μJ mean pulse energy. These pulses were focused onto the sample surface to a spot size of about 20 μm full width at half maximum (FWHM) at normal incidence and the scattered radition was detected in transmission geometry on a suitable area detector (CCD).

Samples comprised 100 nm thick polycrystalline silicon (pc-Si) films, deposited on 100 μm × 100 μm sized $Si_3N_4$ membrane windows (thickness 20 nm) arranged into large arrays supported by a silicon wafer frame. This target design allowed replacement of the sample between consecutive pulses, which was necessary because of damage induced by single-pulse irradiation with either the optical laser or the FEL.

The pc-Si films were irradiated at an angle of incidence of 47° by single 12-ps laser pulses at 523 nm wavelength, focused to an elliptical beam spot of approximately 40 μm × 30 μm in diameter (FWHM). The peak laser fluence was set at ≈1.7 J/cm$^2$ sufficient to completely ablate the silicon film (ablation threshold fluence ≈0.5 J/cm$^2$).

Figure 4(a) compiles a sequence of transient scattering patterns (frames) for six selected pump-probe delay times between −10 ps and +4.5 ns, as indicated in the upper left corner. The intensity scale of the false-color representation is in arbitrary units but identical for all delays/frames. The vertical white bar in frame (a) indicates the projection of the p-polarized pump laser beam polarization onto the sample surface.

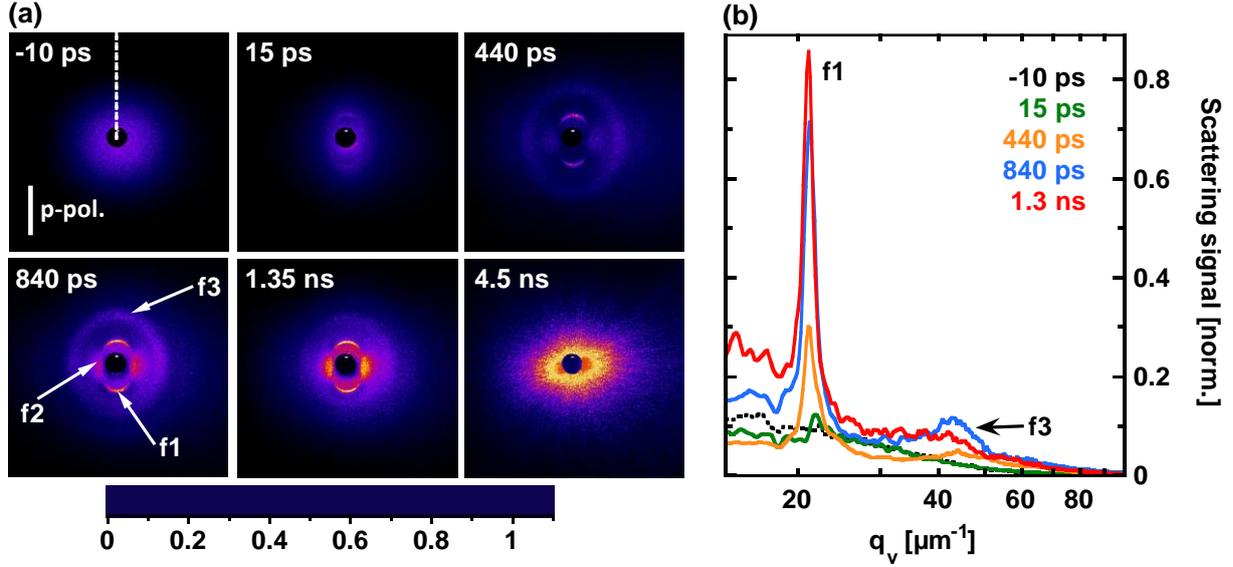

**Fig. 4:** (a) Transient scattering patterns (false-color representation) of a 100 nm thick polycrystalline Si-film as a function of delay time $\Delta t$ between the 12 ps, 523 nm optical pump pulse (peak fluence ≈ 1.7 J/cm$^2$), and the 10-20 fs, 13.5 nm XUV probe pulse. The displayed range of spatial frequencies $q_{h/v}$ (in horizontal and vertical directions) is between −92 μm$^{-1}$ and +92 μm$^{-1}$. The normalized false-color intensity scale is the same for all frames. The solid white bar in the image for $\Delta t$ = −10 ps indicates the laser polarization (p-pol., projection on to the surface). (b): Vertical cross sections along the white dashed line in the image for $\Delta t$ = −10 ps in (a) for different pump-probe time delays.

The scattering pattern measured at negative delay time represents the non-irradiated polycrystalline material. At positive delay times the scattering patterns rapidly develop a rich structure and three characteristic features (labelled as f1 – f3 in Fig. 4) can be identified. Already during the laser pulse the most prominent feature (f1) becomes visible, which exhibits a double-sickle shape (see frame $\Delta t$ = +15 ps). Its intensity increases with time, reaches a maximum at about 1 ns, but it persists until the whole sample gets destroyed (see frame for $\Delta t$ = 4.5 ns). This can be also seen in the vertical cross sections diplayed in Fig. 4(b), where f1 corresponds to the sharp maximum at $q_v$ ≈ 21 μm$^{-1}$, indicative of the formation of a well-oriented structure (perpendicular to the laser polarization) with a spatial periodicity of $\Lambda_{f1}$ ≈ 300 nm. Almost as fast as f1, feature f3 emerges, a broad, slightly elliptical diffraction ring with initial half axes of $q_v$ = 44.5 μm$^{-1}$ and $q_h$ = 37 μm$^{-1}$ in vertical and horizontal direction, respectively. While the integrated scattered signal amplitude of the f3-feature increases until $\Delta t$ ~0.5 ns before falling offand vanishing after about 1 ns, its size in momentum space exhibits a slight but continuous decrease (see Fig. 4(b)). In contrast to f1, the elliptical shape of f3 indicates the formation of randomly oriented periodic structures with periodicities varying from approximately 140 – 150 nm in vertical and 170 – 180 nm in horizontal direction. Feature f2, two broad diffraction maxima at $q_h$ ≈ ±(15−25) μm$^{-1}$ in a scattering direction perpendicular to the laser polarization, develops after approximately 300 ps and increases in intensity till the eventual irreversible disintegration of the sample after a few ns.

## 2.2 Proof of Sipe's LIPSS theory and beyond

As already mentioned above, LSFL originate from a periodic spatial modulation of the energy deposited into the material by optical excitation, which is caused by the interference of surface scattered/excited waves with the incident laser pulse. The first-principles LIPSS-theory by Sipe and co-workers [34] describes this quantitatively through the so-called *efficacy factor* $\eta(q_v, q_h)$, the 2D Fourier-representantion of the spatially varying energy deposition pattern. Under the assumption that this pattern is transformed into a correponding modulation of the surface topography, for example by local ablation of material, the measured scattering patterns should directly reflect the efficacy factor $\eta$.

In Figure 5 we compare, therefore, experimental scattering patterns (top row), measured again on 100 nm thick, polycrystalline Si films, but this time deposited on smaller 20 μm × 20 μm Si$_3$N$_4$-membrane windows and at a delay time of $\Delta t$ = +340 ps, to calculations of $\eta$ (bottom row) using the Sipe-model for p-polarized (left column) and s-polarized (right column) pump pulses. The bright horizontal and vertical streaks in the experimental patterns are caused by scattering of radiation in the spatial wings of the XUV probe-beam at the window edges because of the smaller window size. The black, yellow, and green dashed lines in (a) and (b) mark the feature f1, f2, and f3, respectively, as discussed above. The overall scattering intensity for s-polarized probe pulses is significantly weaker than in the p-polarized case. As a consequence the f2-feature is not visible and "buried" in the strong scattering from the window edges [Fig. 5(b)].

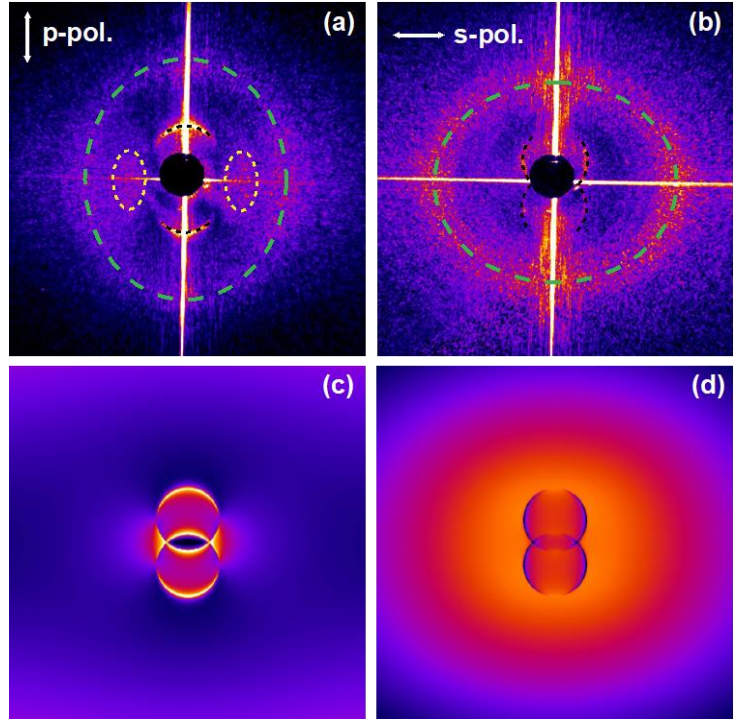

**Fig. 5:** Top row: Transient scattering patterns of a 100 nm polycrystalline Si film (on a 20 μm × 20 μm, 20-nm-thick $Si_3N_4$ membrane) recorded 340 ps after excitation with a 12 ps, 523 nm laser pulse (peak fluence ≈ 2 – 2.2 J/cm$^2$) for two orthogonal polarization directions [(a): p-pol.; (b): s-pol.]. Spatial frequencies $q_{h/v}$ ranging between −70 μm$^{-1}$ and +70 μm$^{-1}$. The black, yellow, and green dashed curves mark the characteristic diffraction features f1, f2, and f3, respectively. Please note that the intensity scale has been individually adjusted to make the different diffraction features clearly visible. The overall scattering intensity is decreasing by changing the pump polarization from p to s. Bottom row: Corresponding efficacy factor ($\eta$) maps [(c): p-pol.; (d): s-pol.] encoding the energy deposition to a rough fully molten (l-Si) surface and standard roughness parameters (shape factor $s = 0.4$, filling factor $F = 0.1$, for details of calculations see [64]).

Clearly all features rotate with the pump polarization, giving evidence that they are all related/influenced by the periodic modulation of the initial lateral energy deposition pattern. Moreover, by comparing the experimental patterns (a, b) to the calculated ones (c, d), the experimental f1-feature can be unambiguously associated with the formation of LSFL [65], as predicted by the LIPSS-theory of Sipe and coworkers [34]. As discussed in more detail in [60], for a p-polarized pump a comparison of vertical profiles along $q_v$ reveals almost perfect quantitative agreement between the experimental data and $\eta$ for later delay times when the f3-feature has already vanished – a convincing proof of the analytical Sipe-theory under *true single-pulse irradiation conditions*.

For an s-polarized pump beam (right column in Fig. 5) the situation is a bit more complicated. The efficacy factor $\eta$ calculated for s-polarization exhibits – as the experimental data – the f1-feature (this time double-sickles oriented along $q_h$), but interestingly it corresponds to spatial frequencies where the energy deposition is *minimal*. Of course, such a grating of reduced modification/ablation leads as well to a positive scattering signal – as the grating of enhanced ablation in the case for p-polarized excitation. So the observed behavior for s-polarized excitation shows also good agreement with the Sipe-theory.

It should also be noted that this excellent agreement could only be achieved by using the optical constants of liquid Si as input for the Sipe-model. This can be rationalized by considering that due to the rather long pump pulse duration and the high fluences discussed here, the laser-excited Si melts already during the optical pump pulse, turning its optical properties from that of a semiconductor (pc-Si) to that of a plasmonically active metal (l-Si).

However, the comparison of the experimental data and the calculated efficacy factor $\eta$, as depicted in Fig. 5, also clearly reveals that the electromagnetic Sipe-theory can neither account for the other features f2 and f3, nor for the complex temporal evolution of the scattering patterns. The latter is actually not surprising since the Sipe-model is a *static* approach, which does not consider transient changes of material properties or any structural dynamics due to heating, melting and ablation.

For example, given its delayed temporal appearance and almost circular symmetry, we attribute f3 to the excitation of highly damped capillary waves on the molten Si surface [60] (possibly in spatial regions in the vicinity of the ablation threshold). While the specific spatial frequencies at which these capillary waves are excited are related to the initial optical energy input, as evidenced by the polarization dependence of the f3-feature, such hydrodynamic effects are not part of the Sipe-model. Similarly, it must be concluded that also the f2-feature may be seeded by the spatial modulation of the optical energy input. However, its delayed appearance and temporal evolution clearly points towards another material-specific effect that still has to be explained.

In summary, the examples discussed in this section provide clear experimental evidence that both electromagnetic (e.g. plasmonic optical excitation) and structural/hydrodynamic effects (e.g. phase transitions, capillary waves, ablation, etc.) determine the formation of nanoscale structures at laser-irradiated surfaces. Through time-resolved scattering measurements with sub-µm/nm spatial and sub-ps temporal resolution these processes can be experimentally revealed and distinguished in time annd momentum space. Moreover, these experiments currently represent the "world record" in resolving small-scale LIPSS structures at high-resolution in space and time.

## 3. Future perspectives

While the results presented in section 2 have revealed a complex evolution of the irradiated material on different length- and time-scales, these experiments are in fact not much more than a proof-of-concept since more detailed investigations (e.g. variation of the pump laser fluence) had not been possible due technical restrictions and the limited amount of available beamtime. Nevertheless, we believe that these experiments have laid the foundation for establishing this technique as a general method for studying structure formation on laser-irradiated surfaces (including LIPSS) on the relevant time and length scales, and some of the possibilities available now or in the near future will be briefly outlined below.

*3.1 XFEL technology development*

Since our experiments have been performed (2007!) the XFEL- as well as the associated beamline technology has seen dramatic developments allowing significant improvements for time-resolved scattering experiments. Almost "trivial" is their extension towards shorter probing wavelengths, which are available at hard X-ray FELs. This will enable experiments with even higher spatial resolution and, therefore, the investigation of the formation of HSFL [as shown in Fig. 1(b)] and other deep sub-µm structures.

Moreover, sophisticated X-ray dectectors have been developed, which combine large size, high sensitivity and dynamic range, and fast readout rates. These detectors not only allow to access a much larger momentum transfer range, but also to record data with higher signal-to-noise and to perform the experiments with higher speed, thus increasing throughput/efficiency.

With respect to future time-resolved studies of LIPSS-formation we would like to discuss as a particular example the in our view unique capabilities of the Spectroscopy and Coherent Scattering (SCS) instrument at the European XFEL in Hamburg, Germany [66 - 68]. SCS uses radiation from the SASE 3 undulator in the soft to tender X-ray range (depending on the electron energy from 0.25 – 3 keV). The so-called Forward-scattering Fixed Target (FFT) experimental chamber [67] facilitates among other techniques, Small Angle X-ray Scattering (SAXS) experiments in transmission geometry under vacuum on X-ray transmissive solid samples [69 - 72]. It is equipped with a fast sample scanner for samples with a size of up to 50 mm × 50 mm, thus allowing for rapid sample replenishment, which is key for experiments in an irreversible excitation regime. Considering, as for the experiments at FLASH, coated $Si_3N_4$ membrane arrays as sample carrier, a 50 mm × 50 mm Si wafer frame would allow for more than 9000 sample windows of 100 µm × 100 µm size. The FFT-chamber is also equipped with a load-lock system for the storage of 7 sample frames, thus enabling rapid sample exchange without breaking vacuum.

The detector at SCS (DEPFET Sensor with Signal Compression – DSSC) has 1 Mpixel and a sensitive area of more than 500 cm$^2$ arranged in a four-segment windmill configuration [73]. It is mounted on a linear translation stage with 5 m travel range, and the Sample-Detector Distance (SDD) can be chosen over a range from 0.4 m to 5.4 m. In addition, rather rapid changes of the SDD by up to 1.5 m from a given set-point are possible without breaking vacuum and modifying the experiment. This in combination with the possibility to change also the X-ray photon energy over an extended range rather quickly, provides great flexibility to adapt the experimental configuration to specific requirements and gives access to a wide range of length scales from the µm-range at low photon energy and large SDD down to the few-nm range at higher photon energies and small SDD.

To pump the sample a synchronized (jitter < 50 fs) optical laser system is available, which delivers ≤ 50 fs pulses at a fundamental wavelength of 800 nm and with up to 2 mJ pulse energy. This radiation can be frequency converted by second and third harmonic generation to 400 nm and 266 nm, respectively. Moreover, an optical parametric amplifier provides wavelength-tunable pulses over a range from 350 nm to 2.5 µm.

Another very interesting possibility is enabled by the flexible pulse pattern available at EuXFEL. The FEL operates in a bunch-train mode with an inter-train repetition rate of 10 Hz and up to a few hundred pulses per train (at 10 Hz) with a minimum separation of 220 ns. The synchronized laser system operates in the same bunch-train mode, but the intra-train pulse pattern can be separately programmed for the FEL and the optical laser. This would allow for example a very time-efficient investigation of LIPSS-formation in the multi-pulse excitation regime and the role of inter-pulse feedback phenomena [27, 74]. The sample surface will be pre-patterned with a number of ($N$ – 1) "dressing" laser pulses (µs pulse separation), before in the *N*-th pump-probe irradiation event the transient state of matter is interrogated at the desired delay time $\Delta t$.

*3.2 New strategies*

As outlined above, soft to tender X-rays allow to reach down to few nm spatial scales. To bridge the final gap to sub-nm and atomic length scales X-ray pulses of even higher photon energy (shorter wavelengths) are required. In fact, already a number of laser pump - Wide Angle X-ray Scattering (WAXS) probe experiments have been performed in the *irreversible* excitation regime, which where addressing for example laser induced melting in thin films and nanoparticles (e.g. [75 - 77]) or other kind of phase transitions in condensed matter (e.g. [19, 78]).

So far we have been discussing scattering experiments in a rather simple normal-incidene transmission geometry. This requires samples that are transmissive for the probe pulse at the given probe radiation wavelength. Therefore, films with a thickness matched to the corresponding absorption length of the material under study must be used. For example, the absorption length of Si at 13.5 nm wavelength (below the Si L-edge) is nearly 600 nm and the 100 nm films used at FLASH clearly fulfill this requirement. In the keV-range the absorption of Si is even weaker, but also higher Z materials exhibit absorption lengths of a few hundred nm to a few µm (depending on material and photon energy), so that X-ray absorption is not a principal restriction.

Nevertheless, one has to ask the question, whether transmission scattering experiments on thin film samples are able to address (all) the relevant aspects about LIPSS-formation. While our FLASH-experiments clearly showed LIPSS-formation and one may argue that films with a few hundred nm thickness essentially exhibit bulk-like properties, real-life laser processing usually deals with bulk materials.

To study structure formation on the (sub-)µm to nm scale at the surfaces of laser-irradiated bulk materials a grazing incidence reflection geometry needs to be applied. Recently Randolph et al. have demonstrated this method for studying plasma formation and ablation of Ta/$Cu_3$N multilayer structures [79]. Despite the higher complexity of such experiments (e.g. keeping spatial overlap between the pump- and the probe-beam when translating the sample, high sensitivity to beam pointing fluctuations, etc.) this technique should be readily applicable to the case of LIPSS-formation.

Besides "simple" scattering experiments, other approaches might be used to study LIPSS-formation. For example, 3D tomography with multiple probe pulses at different angles of incidence may provide additional volumetric information. Moreover, experimental techniques that take benefit of the very high spatial coherence of the FEL radiation, such as coherent diffraction imaging [62, 76, 77, 80] or photon correlation spectroscopy [80 - 82], might be applicable even under strong excitation conditions.

*3.3 Synergy with advanced modelling*

As already pointed out above and despite its success, the LIPSS-theory by Sipe and co-workers is in principal neither able to account for changes of material properties during laser irradiation nor can it describe the subsequent structural response. Recently, sophisticated hybrid electromagnetic and hydrodynamic continuum models [42, 83] have been developed, to enable a "full" description over different spatial and temporal scales.

Such models calculate the deposition of optical energy to the material on the basis of 3D numerical *Finite-Difference Time-Domain* (FDTD) calculations that are numerically solving Maxwell's equations in almost arbitrary boundary conditions that are considering the surface topography of the irradiated material. In combination with a *Two-Temperature Model* (TTM, [84]), the energy relaxation from the laser-excited electronic system to the materials' lattice as well as phase transitions such as melting can be considered and combined with hydrodynamic models on the basis of the *Navier-Stokes Equations* (NSE) and the *Equation of state* (EOS) to include thermodynamic and thermophysical effects (such as capillarity, rarefaction waves, evaporation etc.). Similarly, hybrid TTM – *Molecular Dynamics* (MD) simulations are able to reveal atomic-scale processes for example during melting and ablation of laser-irradiated materials (e.g. [16, 85, 86]).

The scattering experiments discussed here will provide important benchmark data to test and validate these models. In turn, such modelling is essential for the interpretation of complex experimental scattering patterns (e.g. identifying the processes behind the f2-feature). As such this synergistic approach may better answer the "fundamental" LIPSS-question, why and how laser-irradiation can trigger the transition of a "chaotic" (rough) surface from an aperiodic into a periodic surface structure, but it can also guide the development of improved strategies to tailor and control LIPSS-formation for technological applications.

Beyond LIPSS this approach must be viewed in a much wider context: The joint experimental & modelling methodology suggested here opens a new avenue for studying irreversible laser-matter interactions by bridging and "calibrating" material responses over a wide range of length- and time-scales. This will enable new insights into laser-induced surface strucure formation in general [87] and further stimulate the fields of fundamental physics, material sciences, and applications.

**Acknowledgements**

The experiments at the FLASH FEL, from which we have shown here selected data, were done using the coherent diffractive imaging instrument developed by H. N. Chapman, A. Barty and co-workers, to whom we are particularly indebted. In addition, we gratefully acknowledge the many different contributions of (alphabetical order) S. Bajt, M. J. Bogan, S. Boutet, A. Cavalleri, S. Düsterer, H. Ehrke, M. Frank, J. Hajdu, S. Hau-Riege, S. Marchesini, M. M. Seibert, N. Stojanovic, R. Tobey, R. Treusch, and the late B. Woods. We thank G. Mercurio and A. Scherz for providing detailed information about the SCS-instrument of the European X-ray Free Electron Laser facility. This work was financially supported by the *Deutsche Forschungsgemeinschaft* (DFG, German Research Foundation) through the Collaborative Research Centre (CRC) 1242 (project number 278162697, project C01 *Structural Dynamics in Impulsively Excited Nanostructures*).